# Multi-GeV Electron Combs from a Plasma Wakefield Accelerator

Chaojie Zhang[1,*], Douglas Storey[2,*], Alexander Knetsch[2], Brendan D. O'Shea[2], Robert Ariniello[2], Sébastien Corde[4,2], Thamine N. Dalichaouch[5], Claudio Emma[2], Ole G. Finnerud[3], Spencer Gessner[2], Claire Hansel[6], Valentina Lee[6], Carl A. Lindstrøm[3], Michael Litos[6], Nathan Majernik[2], Kenneth A. Marsh[1], Warren B. Mori[5], Mark J. Hogan[2], and Chan Joshi[1,*]

1. Department of Electrical Engineering, UCLA, Los Angeles, California 90095, USA

2. SLAC National Accelerator Laboratory, Menlo Park, California 94025, USA

3. Department of Physics, University of Oslo, Oslo 0316, Norway

4. Laboratoire d'Optique Appliquée (LOA), CNRS, École polytechnique, ENSTA, Institut Polytechnique de Paris, Palaiseau, France

5. Department of Physics and Astronomy, UCLA, Los Angeles, California 90095, USA

6. Department of Physics, University of Colorado Boulder, Boulder, Colorado 80309, USA

Corresponding authors: chaojiez@ucla.edu, dstorey@slac.stanford.edu, cjoshi@ucla.edu

**Abstract**

Plasma accelerators now produce GeV-class electron beams with brightness and stability sufficient to drive free-electron lasers. Beyond this, they possess a unique yet largely unexplored capability: shaping the phase space of the beam in situ during injection, on femtosecond or shorter timescales. Here we demonstrate this capability by generating a multi-GeV electron comb comprising more than ten microbunches simultaneously separated in both energy and time. Periodic pinching of the drive beam inside its self-excited plasma wake sequentially injects microbunches via ionization of embedded helium atoms at successive betatron oscillations, while the gently varying plasma density maps each bunchlet to a distinct wake phase, compressing electrons trapped over a ~17 cm region into a comb only micrometers long. Individual microbunches exhibit percent-level energy spreads, energy spacing up to ten percent, and contain several picocoulomb charge. The percent-level spreads and parabolic energy-spacing trend provide experimental evidence for sub-femtosecond microbunch durations and few-femtosecond separations as revealed by beam-loading analysis and confirmed by particle-in-cell simulations. This work demonstrates femtosecond, in-situ phase-space shaping in plasma accelerators, paving the way for electron beams with tailored energy-time structure.

Plasma-based accelerators driven by either a relativistic charged particle beam [1–4] or an intense laser pulse [5–8] have demonstrated accelerating gradients exceeding 10 GV/m [9–12] in the nonlinear blowout regime, enabling multi-GeV electron beams in centimeters [12–16]. Plasma-accelerated beams now approach the quality and stability of those from conventional radio-frequency (RF) accelerators [17], and have driven free-electron lasers (FELs) at wavelengths from the infrared to extreme ultraviolet [18–21]. Most recently, a plasma wakefield accelerator has simultaneously boosted beam energy and brightness to levels relevant to hard X-ray FELs and future colliders [16].

These advances have focused on producing beams whose quality matches that from conventional accelerators [17,22–24]. Yet plasma accelerators also possess a unique capability that remains largely unexploited: when electrons are trapped inside the plasma wake to form a new beam [25–29], their phase space can be shaped in situ, at the moment of trapping, on femtosecond or shorter timescales. This contrasts with conventional approaches to generating structured electron beams, which rely on either photocathode pulse shaping [30–32] or post-acceleration beam manipulation using modulators and compressors [33], and which require increasingly elaborate multi-stage setups at shorter timescales [34]. Structured beams are attracting growing attention in accelerator and photon science: microbunched beams can drive high-power coherent radiation sources [30–32], and multichromatic bunch trains could enable single-shot ultrafast stroboscopic measurements [35] and multi-color X-rays [36–38].

Several schemes exploiting this in-situ beam shaping capability have been proposed and explored in particle-in-cell (PIC) simulations. These include generation of multi-energy bunches via dual-color laser-triggered ionization injection [39]; pre-bunched beams at angstrom wavelengths through modulated density downramps [40]; and beams carrying orbital angular momentum via ionization injection using structured light [41,42]. Experimentally, energy-bunched spectra from ionization injection have been observed in beam-driven [43] and laser-driven [44] plasma wakes, the latter arising from subcycle carrier-envelope-phase effects. However, none of these experiments have produced multi-GeV beams with microbunches that are well-separated in both energy and time.

Here we demonstrate such a beam: a multi-GeV electron comb comprising more than ten microbunches with distinct energies and femtosecond separations, generated in a beam-driven plasma wakefield accelerator at the FACET-II facility. The comb arises from density-gradient-mediated phase-space mapping combined with betatron-driven periodic ionization injection.

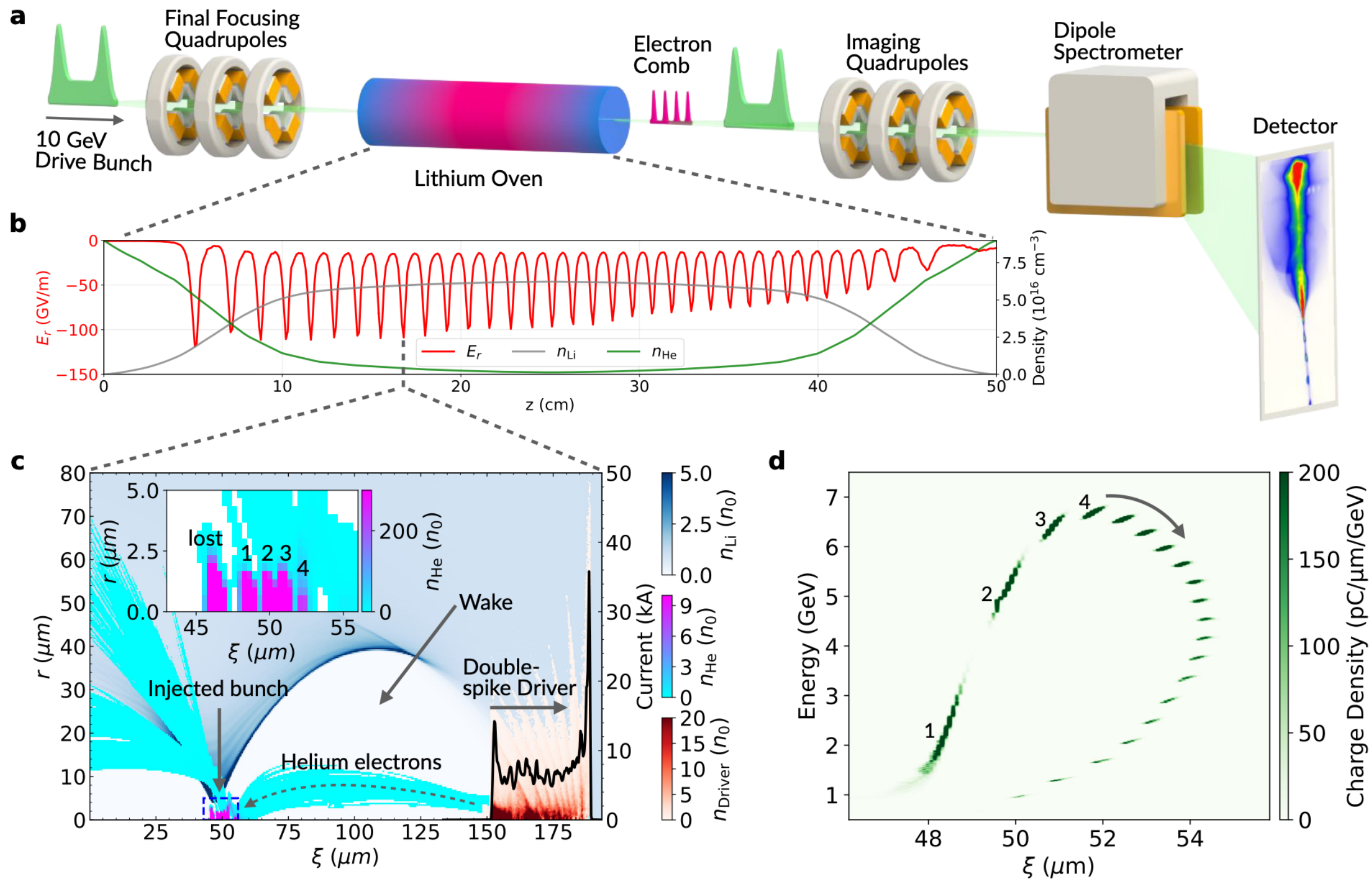

**Figure 1. Electron comb generation in a plasma wakefield accelerator. a**, Experimental layout. A 10-GeV electron drive beam with a double-spike current profile is focused into a lithium heat-pipe oven, where it ionizes lithium, excites a nonlinear plasma wakefield, and injects electrons from helium buffer gas. The resulting electron comb and decelerated drive beam are energy-dispersed by an imaging spectrometer onto scintillator screens. The decelerated drive beam, which acquires a broad continuous energy spread in exciting the wake, appears as a diffuse distribution on the detector. **b,** The expected neutral density profiles of lithium (gray) and helium (green) along the propagation axis, with the transverse electric field of the trailing current spike (red, evaluated at $r = 0.1c\omega_p^{-1}$) showing periodic betatron pinches where ionization injection occurs. **c**, PIC simulation snapshot showing the plasma wake (white ion bubble enclosed by the electron sheath, blue) excited by the double-spiked 10-GeV drive beam (orange), at the location indicated by the dashed line in **b**. Helium electrons (cyan) ionized by the transverse field of the trailing current spike slip backwards in the co-moving frame (dashed curve) and are trapped in the back of the wake and form an electron comb (magenta). Inset: four bunchlets (labeled 1-4) sequentially injected at successive betatron pinches; the leftmost is later lost due to wake contraction in the slowly increasing density. **d**, Longitudinal phase space of the electron comb at plasma exit. Distinct bunchlets appear as localized peaks separated in both energy and time $\xi \equiv z - ct$ (co-moving coordinate). The curved arrow traces injection order, showing that early-injected (1-11) and late-injected bunchlets (12-20) follow a reversed order along $\xi$ due to bidirectional density-gradient-mediated mapping (see text).

The experiment was performed at the FACET-II user facility at SLAC National Accelerator Laboratory. Figure 1a shows the schematic of the experiment. The 10-GeV, 1.6 nC electron beam was over-compressed in the linac to produce a drive beam with two narrow, high-current spikes (one at the front and the other at the very back) and was focused into a lithium heat-pipe oven, where helium buffer gas confines the lithium vapor at both ends (Fig. 1b). The beam drives a nonlinear wake inside the lithium plasma and injects helium electrons to form a comb. Upon exiting the plasma, the decelerated drive beam and the electron comb were captured and dispersed by an imaging spectrometer [45] for spectral characterization.

The physical mechanism of generating multi-GeV electron combs is illustrated in Fig. 1c (PIC simulation using the OSIRIS code [46]). The leading current spike (~35 kA) of the drive beam ionizes lithium vapor to create the plasma, while the subsequent middle part of the beam that contains most of the charge drives the wake by transversely expelling plasma electrons. The transverse field of the ions inside the wake exerts a transverse focusing force on the drive beam that induces energy-dependent transverse envelope (betatron) oscillations that periodically pinch the beam [10]. At each pinch, the trailing current spike is focused to below 5 μm, producing a transverse electric field exceeding ~80 GV/m (red curve in Fig. 1b), which is sufficient to ionize the first electron of helium inside the wake (Fig. 1c, cyan), but not the second electron due to its much higher ionization potential. The leading current spike, propagating at its vacuum spot size (~40 μm), cannot reach this ionization threshold. These helium electrons, released within a narrow wake phase, slip backward relative to the wake and become trapped at the rear of the bubble once they gain sufficient energy to co-propagate at nearly the speed of light. This process repeats every half betatron wavelength ($\lambda_\beta = 2\pi\sqrt{2\gamma_b}c\omega_p^{-1}$, $\lambda_\beta/2 \sim 1.35$ to 1 cm for the rear spike as it decelerates from initially 10 to ultimately ~5 GeV in plasma with skin depth $c\omega_p^{-1} \approx 21.7$ μm) as the beam pinches twice per oscillation, injecting electron bunchlets (the first four are labeled 1-4 in Fig. 1c inset) that are subsequently accelerated to multi-GeV energies (Fig. 1c, magenta). The full dynamics are shown in the Extended Data Movie.

The gentle density variation along the plasma column (Fig. 1b) imprints a distinctive phase-space structure on the injected beam through density-gradient-mediated phase-space mapping. In regions of increasing density (z=10-25 cm), the wake continuously contracts, so newly ionized helium electrons are released closer to the peak of the wake potential (at the center of the bubble) [43]. These electrons thus require less backward slippage to become trapped, and consequently settle at positions ahead of earlier-injected bunchlets—a positive $z-\xi$ mapping. At the density maximum (z≈25 cm), where the gradient vanishes, the mapping slope approaches infinity and several bunchlets can overlap in $\xi$. Beyond this point, the density decreases so the wake expands, and the mapping reverses sign: later-injected bunchlets now settle behind previous ones. The result is extreme longitudinal mapping: in this simulation helium electrons ionized discretely and injected over a region of ~20 cm are mapped into an accelerated bunch just ~6 μm long, yielding a mapping factor exceeding 30,000.

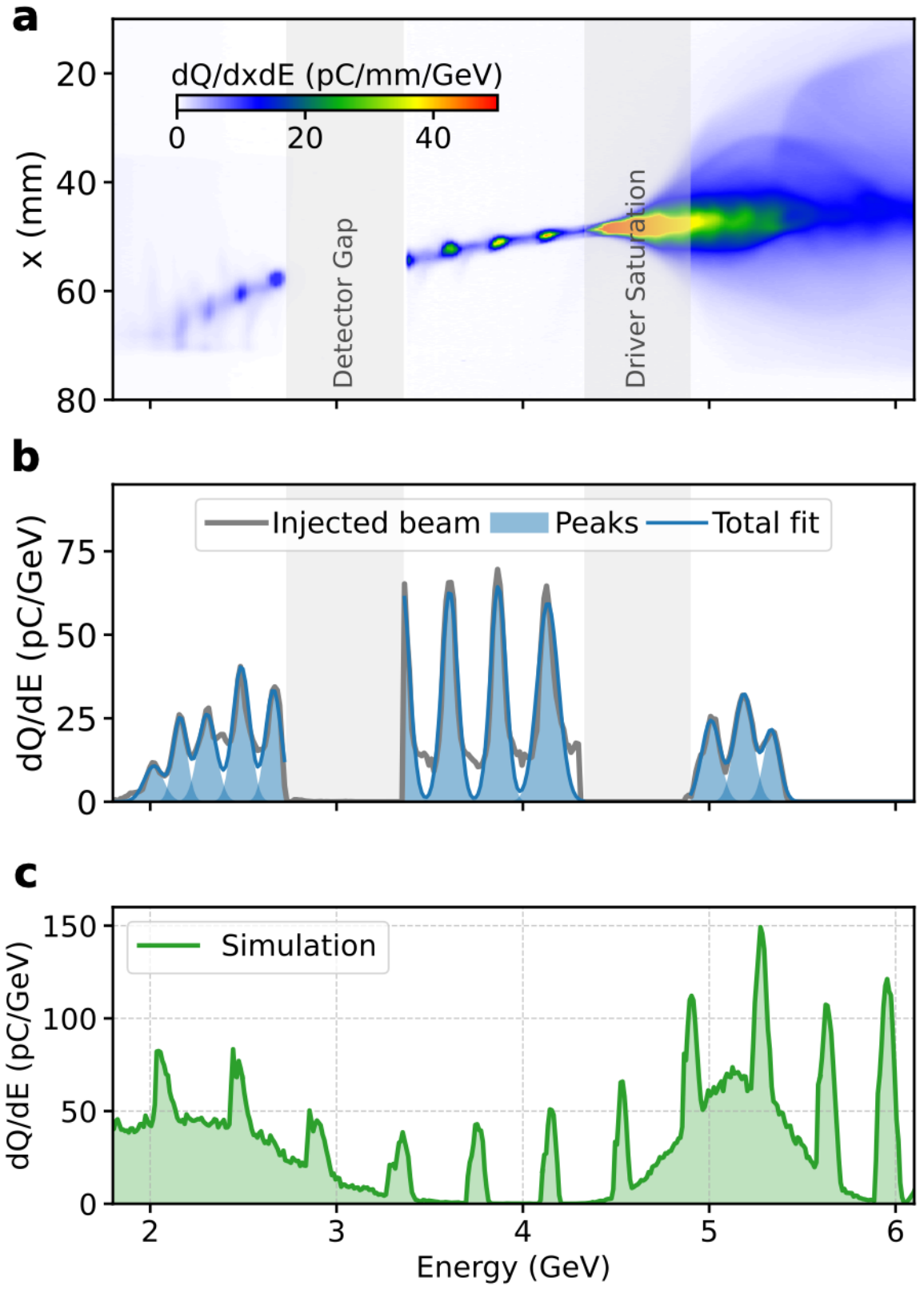

**Figure 2. Experimental observation of the electron comb. a**, Single-shot energy-dispersed image showing the injected electron comb and decelerated drive beam (the transversely broad distribution at higher energies). Gray bands indicate a detector gap (2.8-3.3 GeV) and driver saturation region (4.3-4.9 GeV). **b**, Energy spectrum integrated over the non-dispersive direction (gray) with simultaneous Gaussian fits to individual peaks (blue shaded) and their sum (blue line). Twelve peaks are identified, with spectral charge density up to ~70 pC/GeV. **c**, Simulated energy spectrum (projection of the phase space in Fig. 1d onto the energy axis within 2-6 GeV) for comparison, showing qualitative agreement with the experiment: both exhibit more than ten peaks with deep modulation.

Figure 1d shows the longitudinal phase space of the injected beam at plasma exit: distinct bunchlets appear as localized peaks separated in both energy and time (the co-moving coordinate maps directly to time in the ultra-relativistic beam frame). The curved arrow traces the injection order, revealing the bidirectional mapping. The energy spectrum of the electron comb exhibits more than ten narrow peaks spanning 1-7 GeV, a spectral signature directly accessible to the imaging spectrometer described below.

Figure 2a shows a single-shot spectrometer image of the injected electron comb: a series of transversely narrow, quasi-monoenergetic peaks spanning 1.8-5.5 GeV. The decelerated drive beam appears as a broad distribution above ~5 GeV. Two gray bands mark regions excluded from analysis: a physical gap between spectrometer screens (2.8-3.3 GeV) and a region where drive-beam saturation obscures the signal (4.3-4.9 GeV). The injected beam is transversely much narrower than the driver, even at energies far from the quadrupole focus at 4.3 GeV. This indicates mm-mrad-level emittance consistent with ionization injection, which is much smaller than that of the drive beam (see Methods and Extended Data Fig. 4).

The integrated spectrum (Fig. 2b) quantifies the comb structure. Twelve peaks are identified, each well fitted by a Gaussian distribution. Their rms widths are 46±6 MeV, corresponding to relative energy spreads of 0.8% to 2.6%, with spectral charge density reaching ~70 pC/GeV near 4 GeV. Each peak contains several picocoulombs of charge. The modulation depth approaches 0.8, with charge density between peaks dropping to ~10 pC/GeV. This deep modulation provides strong evidence that the spectral peaks correspond to physically distinct microbunches, rather than to energy modulation of a continuous beam by wake deceleration combined with betatron scalloping. The simulation (Fig. 2c) reproduces the key features: more than ten deeply modulated peaks spanning a comparable energy range with similar spacing.

The comb structure emerges only under specific focusing conditions of the drive beam. Scanning the drive beam's vacuum waist position relative to the plasma entrance (see Extended Data Fig. 1) reveals a clear progression. Far upstream (z=-65 cm; z is measured from the plasma entrance, see Fig. 1b), only drive-beam deceleration is observed. As the waist moves toward the plasma (z=-40 cm), the beam drives small-amplitude wakes and ionizes a small amount of helium atoms, producing low-charge bunchlets that underload the wake and yield broad energy spreads with weak spectral modulation. Near the optimum (z≈10 cm), well-separated peaks with deep modulation appear. Moving the waist further downstream reverses this progression symmetrically, demonstrating experimental control over comb formation.

The multi-peaked structure was observed in approximately 20% of 100 consecutive shots at the optimal waist position. Shot-to-shot variations arise primarily from fluctuations in the drive-beam current profile caused by radio-frequency jitter in the upstream linac [47], a known limitation when delivering ultrashort beams at FACET-II. Extended Data Fig. 2 shows an additional representative shot and a gallery of twenty spectra, all exhibiting comb-like structure with varying peak positions and magnitudes. Simulations with varied parameters—pre-ionized versus beam-ionized lithium, increased current spike spacing, and reduced helium density—all produce comb structures, confirming the robustness of the mechanism.

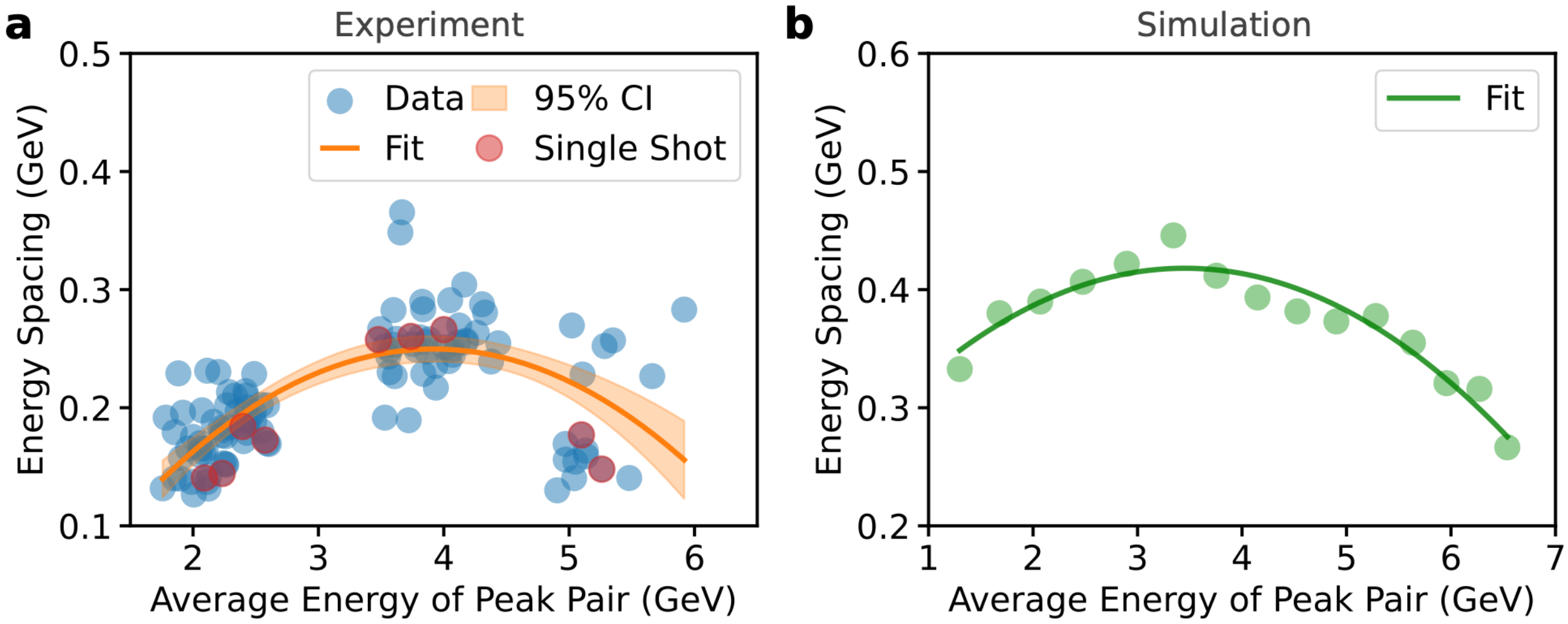

**Figure 3. Energy spacing informs temporal structure of the electron comb. a**, Energy spacing between adjacent peaks versus their mean energy for all comb shots (blue; see Extended Data Fig. 2). Red dots: single shot from Fig. 2. Orange curve and shaded band: parabolic fit with 95% confidence interval, showing maximum spacing of 0.26±0.03 GeV near 4 GeV. The parabolic trend—reduced spacing at both low and high energies relative to the maximum—reflects beam loading by early-injected bunchlets (see text). **b**, Simulated energy spacing, reproducing the parabolic trend with a maximum of ~0.4 GeV near 4 GeV.

The spectral characteristics of the electron comb inform its temporal structure. Sequential betatron-induced ionization injection, combined with phase-space mapping mediated by the plasma density gradient, inherently produces temporally separated bunchlets (Fig. 1d). All measured bunchlets exhibit percent-level relative energy spreads (mean 1.3%, standard deviation 0.5%), with simulations reproducing this range (Fig. 2c). If all bunchlets overlapped in time, later-injected electrons would accumulate charge at the same wake phase, causing cumulative beam loading [48]. In that scenario, only bunchlets experiencing optimal loading would achieve small energy spreads; the remainder would experience either under-loaded or over-loaded wakefield and exhibit larger spreads. The percent-level energy spreads observed across all measured bunchlets thus support temporal separation.

Figure 3a shows that the energy spacing between adjacent bunchlets increases from ~0.15 GeV at 2 GeV to a maximum of 0.26±0.03 GeV near 4 GeV, then decreases toward higher energies—the trend is parabolic. The spacing is set by the accelerating gradient experienced over one half-betatron wavelength (the distance between successive injection points). This parabolic trend constrains individual bunchlet durations. Early-injected bunchlets traverse the longest acceleration interval between successive injections (due to the higher drive beam energy and lower plasma density) and reside closest to the bubble rear, where the gradient is steepest. Without significant beam loading, these bunchlets would reach the highest energy and exhibit the largest energy spacing. The observed reduction in spacing at high energies therefore demonstrates that early-injected bunchlets carry sufficient current to substantially reduce the local accelerating gradient by loading the wake. Simulations with reduced helium density, which produce weaker beam loading, yield a monotonic rather than parabolic spacing trend (Extended Data Fig. 3), confirming this interpretation. Significant wake modification requires peak currents of several kA [48]; with picocoulomb-level charge per bunchlet, this implies sub-femtosecond durations, consistent with simulations (Fig. 1d, Extended Data Fig. 3c).

The measured energy spacing also allows estimation of the beam length and microbunch separations. The microbunches with the largest energy spacing are injected near the middle of the plasma where the drive beam has lost substantial energy. Since the minimum observed driver energy is 4-6 GeV (Fig. 2, Extended Data Figs. 1 and 2), the trailing current spike energy at mid-plasma is approximately 7-8 GeV, giving a half betatron wavelength of ~1.2 cm. The maximum energy spacing of 0.26 GeV then corresponds to a loaded accelerating gradient of 22 GV/m, or approximately $0.94E_0$, where $E_0 \equiv m_e c\omega_p e^{-1} \approx 24$ GV/m is the cold-plasma wavebreaking limit. In the blowout regime, the accelerating field increases linearly from the wake center towards the rear, with a slope of ~1/3 $E_0$ per skin depth for our intermediate ($r_b \approx 2c\omega_p^{-1}$) blowout radius [49]. The measured gradient thus indicates that the injected bunch occupies $\sim 0.3c/\omega_p$, or 7 μm, at the rear of the bubble. This gives a few-femtosecond separation between the twelve resolved microbunches. Taking into account additional microbunches obscured by the detector gap and driver saturation region (Fig. 2), the injection region spans approximately 17 cm, yielding a mapping factor of ~26,000, consistent with the value of ~30,000 (see Extended Data Fig. 5) obtained from simulations.

Together, the percent-level energy spreads and energy spacing provide experimental evidence that the electron comb comprises temporally distinct sub-femtosecond microbunches separated by a few femtoseconds. Two features of the present configuration enable this temporal bunching: the short trailing current spike ionizes helium within a narrow wake phase at each betatron pinch, producing sub-femtosecond trapped bunchlets, while the gentle density gradient maps sequentially injected bunchlets to distinct $\xi$ positions, ensuring their temporal separation. This contrasts with configurations using longer Gaussian drive beams in uniform plasmas, where ionization spans a broader wake phase, producing longer bunchlets that overlap temporally [43].

In summary, we have demonstrated in-situ beam phase space shaping in plasma accelerators, producing the first multi-GeV electron comb. By combining betatron-induced sequential ionization injection with density-gradient-mediated mapping, we compress electrons ionized over 17 cm into a 7 μm beam, corresponding to a mapping factor of approximately 26,000. The comb typically comprises more than ten microbunches spanning 2-6 GeV. The percent-level energy spreads of the microbunches and the parabolic spacing trend reveal that the comb comprises sub-femtosecond microbunches separated by a few femtoseconds, as confirmed by PIC simulations.

The phase-space mapping mechanism demonstrated here, where mapping factor and sign are set by local density gradients, paves the way for generating electron beams with tailored energy-time chirps. By adjusting the plasma density profile and drive beam parameters, the same mechanism could produce combs with narrower total energy span, better suited for driving multi-color XFELs. Moreover, similar mapping should occur in modulated downramp injection schemes. Modulating the downramp at optical wavelengths could produce beams pre-bunched at X-ray wavelengths for direct XFEL seeding. The large energy chirp intrinsic to these combs also opens a path toward attosecond pulse generation via external magnetic compression.

## Methods

### A. Plasma Source

The lithium heat-pipe oven operates at approximately 800°C, producing a lithium vapor column with a neutral density of ~$6\times10^{16}$ $cm^{-3}$ in the near-plateau region. Helium buffer gas at both ends confines the lithium vapor through collisions, creating ~10 cm density ramps at the entrance and exit (Fig. 1b). The central ~30 cm region exhibits gentle gradients with a density maximum near z=25 cm. The density profiles were constructed from on-axis temperature measurements performed ex situ before the experiments; actual profiles during operation may differ slightly due to temperature variations. The much higher ionization potential of helium (24.6 eV) compared with lithium (5.4 eV) ensures that helium remains neutral when the leading current spike ionizes lithium. Only at betatron pinches, where the trailing spike focuses to a few-micron spot size and its transverse field peaks, is helium ionized.

### B. Drive Beam

The FACET-II linear accelerator was configured to over-compress the 10-GeV, 1.6 nC electron bunch, producing a double-spike current profile. An X-band transverse deflecting cavity (XTCAV) measured an overall bunch length of ~20 μm (rms), though this single value does not capture the complex, non-Gaussian current profile revealed by start-to-end beamline simulations. Based on the beamline simulation, we adopted the double-spike current profile shown in Fig. 1c: a leading spike (~35 kA, ~2 μm FWHM) and a trailing spike (~15 kA, ~3 μm FWHM) separated by approximately 35 μm. A final-focusing quadrupole triplet focused the drive beam to a spot size of approximately 38x41 μm (rms) with a beta function of 50 cm at the vacuum waist. The vacuum waist position was scanned by adjusting the final-focusing quadrupole strengths while maintaining a constant beta function. This waist scan demonstrated experimental control over comb formation, with well-separated, narrow spectral peaks emerging only near the optimal waist position (Extended Data Fig. 1).

### C. Energy Spectrometer

The imaging spectrometer comprises a quadrupole triplet for beam capture and refocusing, a dipole magnet for vertical energy dispersion, and two DRZ-FINE scintillator screens imaged by cameras. The first screen, placed in vacuum ~15 m downstream of the plasma exit, covered the low-energy range (1-2.8 GeV). The second screen, placed in air ~5 m farther downstream, covered the high-energy range (3.3-11 GeV). This dual-screen configuration enabled simultaneous measurement over a broad energy range, with a 2.8-3.3 GeV gap. Energy calibration was performed by scanning the dipole field strength and tracking the centroid position of the unperturbed 10 GeV drive beam, achieving better than 1% uncertainty. The energy resolution was approximately 0.15% within the range used, sufficient to resolve the observed comb structures. Absolute charge calibration was performed before data-taking by correlating scintillator signals from non-interacting drive beams to upstream toroid measurements. The spectrometer quadrupoles were set to focus ~4.3 GeV electrons onto the second screen, causing driver signal saturation in the ~4-5 GeV region where decelerated drive electrons overlap with the injected beam. The signal within this range was thus excluded from data analysis. The dispersed spectrum shows a transverse tilt due to slight beam misalignment in the non-dispersive plane at the spectrometer entrance.

**D. Data Analysis**

Raw spectrometer images were processed to produce linearized energy spectra. The energy-dispersion axis of each image was converted to a linear energy scale using the calibrated dispersion relation of the spectrometer. One-dimensional energy spectra were generated by subtracting background and integrating along the non-dispersive axis. In the energy range where decelerated drive-beam and injected-beam signals overlap but the detector is not saturated (above the "Driver Saturation" band in Fig. 2a), the injected beam signal was extracted by fitting a low-order polynomial to the smooth drive-beam baseline and subtracting it. The drive beam contribution forms a smooth baseline, whereas the injected beam produces narrow peaks sitting on top of it, making the two components distinguishable. Peaks in this overlap region were excluded from charge and energy spread calculations but retained for energy spacing analysis, as the baseline subtraction negligibly affects peak energy identification.

Shots exhibiting comb-like structure were identified by the presence of multiple distinct peaks with clear modulation in the 1.8-6 GeV range. Approximately 20% of 100 consecutive shots at the optimal waist position met this criterion. Extended Data Fig. 2 presents an additional representative shot and a gallery of twenty selected spectra showing comb structure with varying peak positions and magnitudes. For qualifying shots, a Gaussian fitting algorithm simultaneously extracted the central energy, rms energy spread, and charge for each peak.

The emittance of the injected beam was estimated from the transverse beam size in the non-dispersive direction on the spectrometer screen. The resulting upper bound, 20 mm-mrad, is set by the detector's spatial resolution. Recent measurements of downramp-injected beams using the same spectrometer and a high-resolution detector showed actual emittances as low as 2 mm-mrad despite apparent values of tens of mm-mrad on the standard detector used in this experiment [16]. The ionization-injected beam in this work is expected to have comparable mm-mrad-level emittance, consistent with the simulated values in Extended Data Fig. 4.

**E. Particle-in-Cell Simulations**

Simulations were performed using the code OSIRIS (version 4.0) in cylindrical (r-z) geometry with a moving window of dimensions 193 μm (z) by 118 μm (r), grid resolution of 0.34 μm (1/64 $c\omega_p^{-1}$) in both directions, and a 0.56 fs time step (1/128 $\omega_p^{-1}$). This resolution is sufficient to resolve the wake structure and the comb formation dynamics, though the internal structure of individual bunchlets approaches the grid scale. Ion motion was not included. A customized second-order field solver was used to mitigate numerical errors and suppress numerical Cherenkov instability. To ensure numerical fidelity over the 50-cm simulation length (~3 million time steps), lithium was initialized as pre-ionized plasma to avoid accumulation of numerical noise at the ionization front. Helium remained neutral and was ionized using the ADK tunnel ionization model [50]. The lithium plasma and helium gas used the density profiles shown in Fig. 1b. The lithium plasma, helium electrons, and drive beam were represented by 16, 16, and 4 macroparticles per cell, respectively.

The 10 GeV drive beam was initialized with the double-spike current profile described above, with a 40 μm (rms) transverse spot size, 40 mm-mrad normalized emittance, and a beta function of 50 cm. The total charge was reduced to 1.0 nC (from the nominal 1.6 nC) to account for imperfect focusing, evidenced by the measured spot size (~40 μm) exceeding the 28 μm expected from the 50 cm beta function and 30 mm-mrad emittance. The remaining charge resides in a low-density halo that contributes negligibly to wake excitation.

Particle tracking was enabled for helium electrons to perform the $z$-$\xi$ mapping analysis presented in Extended Data Fig. 5. Simulations with varied parameters—including beam-ionized lithium (rather than pre-ionized), increased current spike spacing (42 μm), and reduced helium density (5 times lower)—all produced similar comb structures, confirming the robustness of the generation mechanism.

### Extended Data Figures

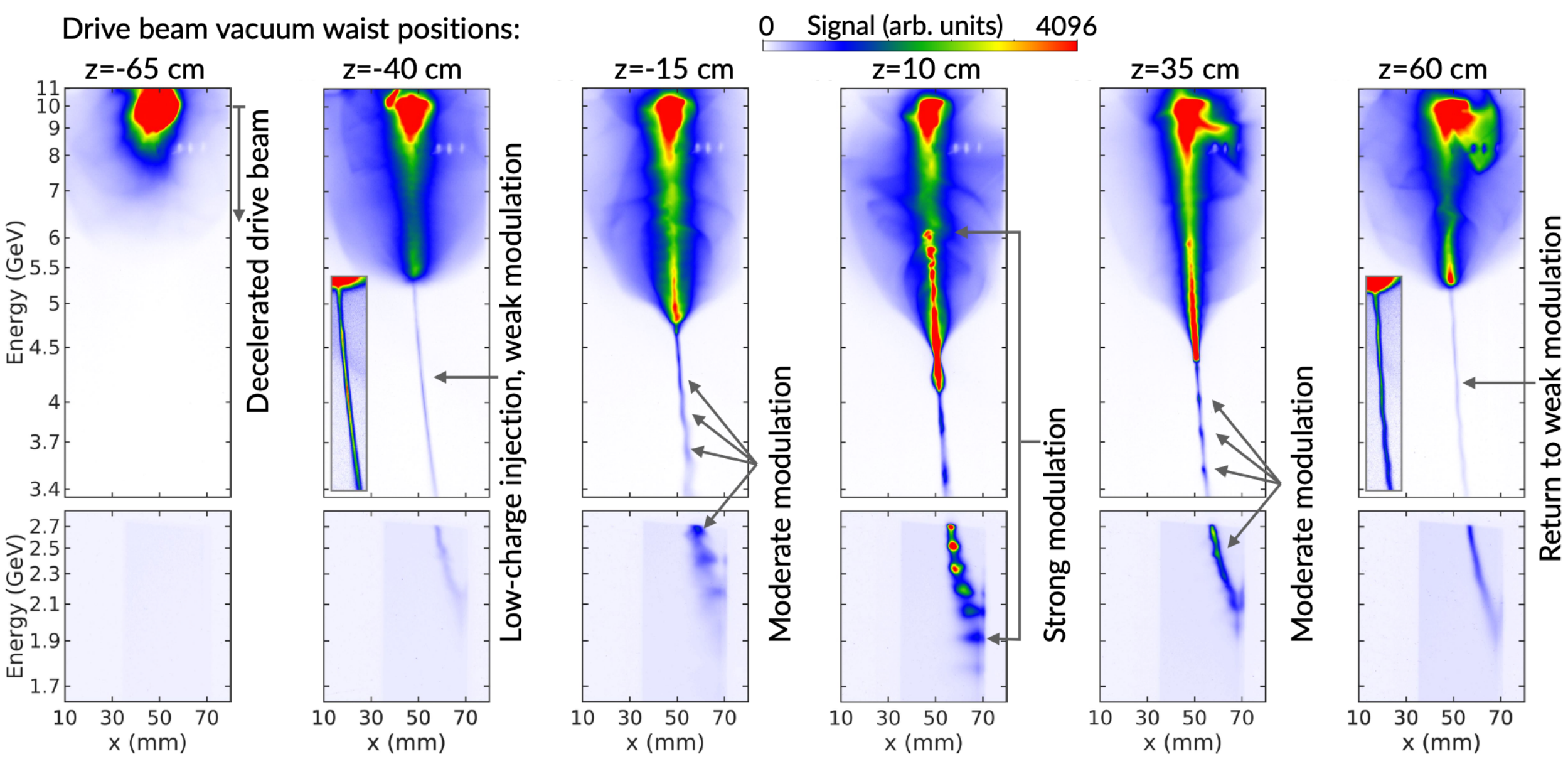

**Extended Data Fig. 1. Waist scan demonstrating control over comb formation.** Single-shot energy spectra for six drive beam vacuum waist positions relative to the plasma entrance. Upper panels: high-energy spectrometer screen (3.4-11 GeV); lower panels: low-energy screen (1.7-2.8 GeV). At z=-65 cm, only drive beam deceleration is observed. As the waist approaches the plasma (z=-40 cm), the drive beam starts exciting small-amplitude wakes and injecting helium electrons, yielding low-charge bunchlets that underload the wake. This leads to larger energy spreads and weaker spectral modulation (inset; color scale reduced by 10x). Near optimal focusing (z=10 cm), higher-charge bunchlets produce strong modulation with well-separated, narrow peaks. The progression reverses symmetrically as the waist moves downstream (z=35 cm, 60 cm).

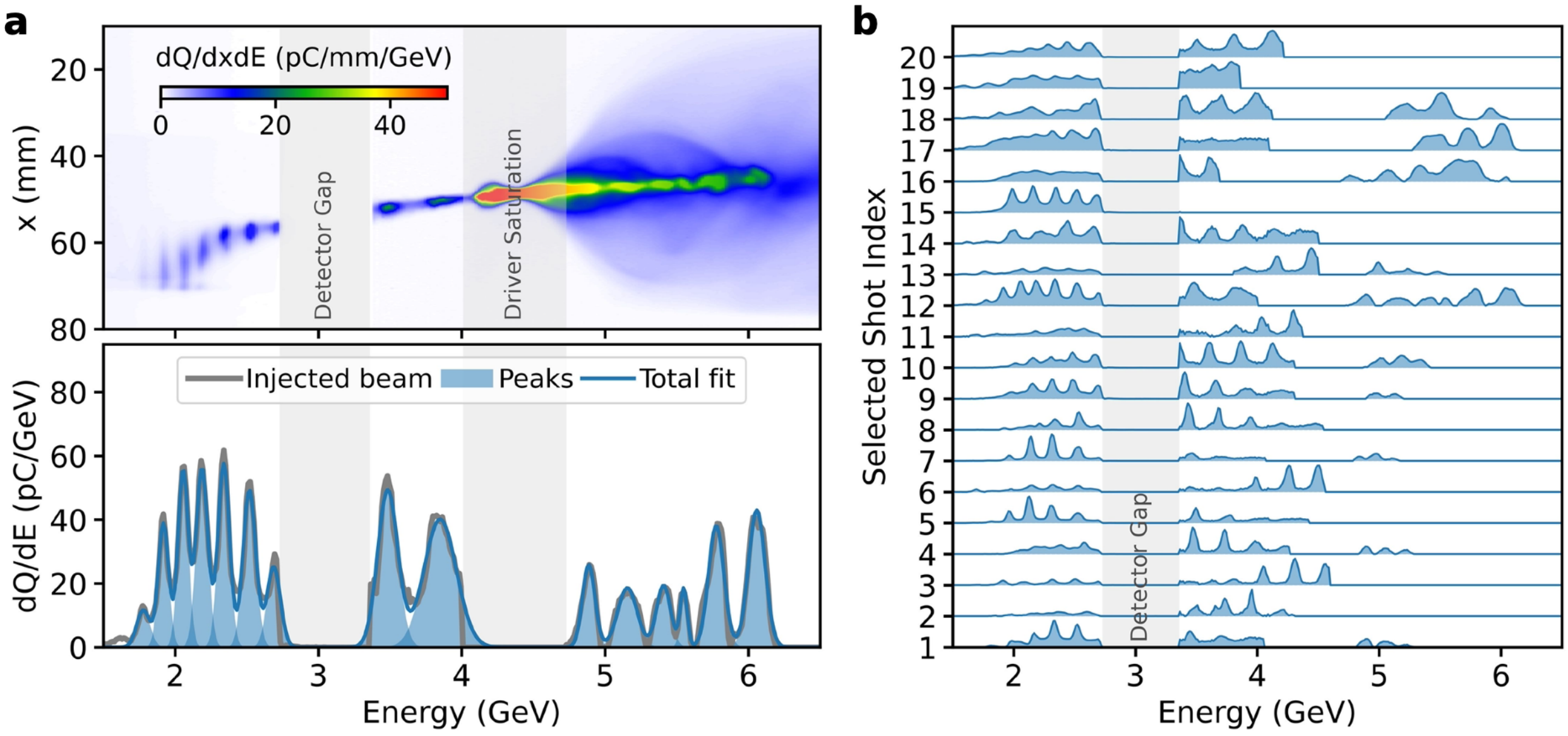

**Extended Data Fig. 2. Experimental reproducibility of the comb structure. a**, Additional representative shot showing energy spacing up to 0.37 GeV near 3.7 GeV. Top: energy-dispersed spectrometer image; bottom: integrated spectrum with Gaussian fits. **b**, Gallery of twenty selected shots exhibiting comb-like structure, demonstrating reproducibility despite shot-to-shot variations in peak positions and magnitudes. The gray band indicates the physical gap between detectors.

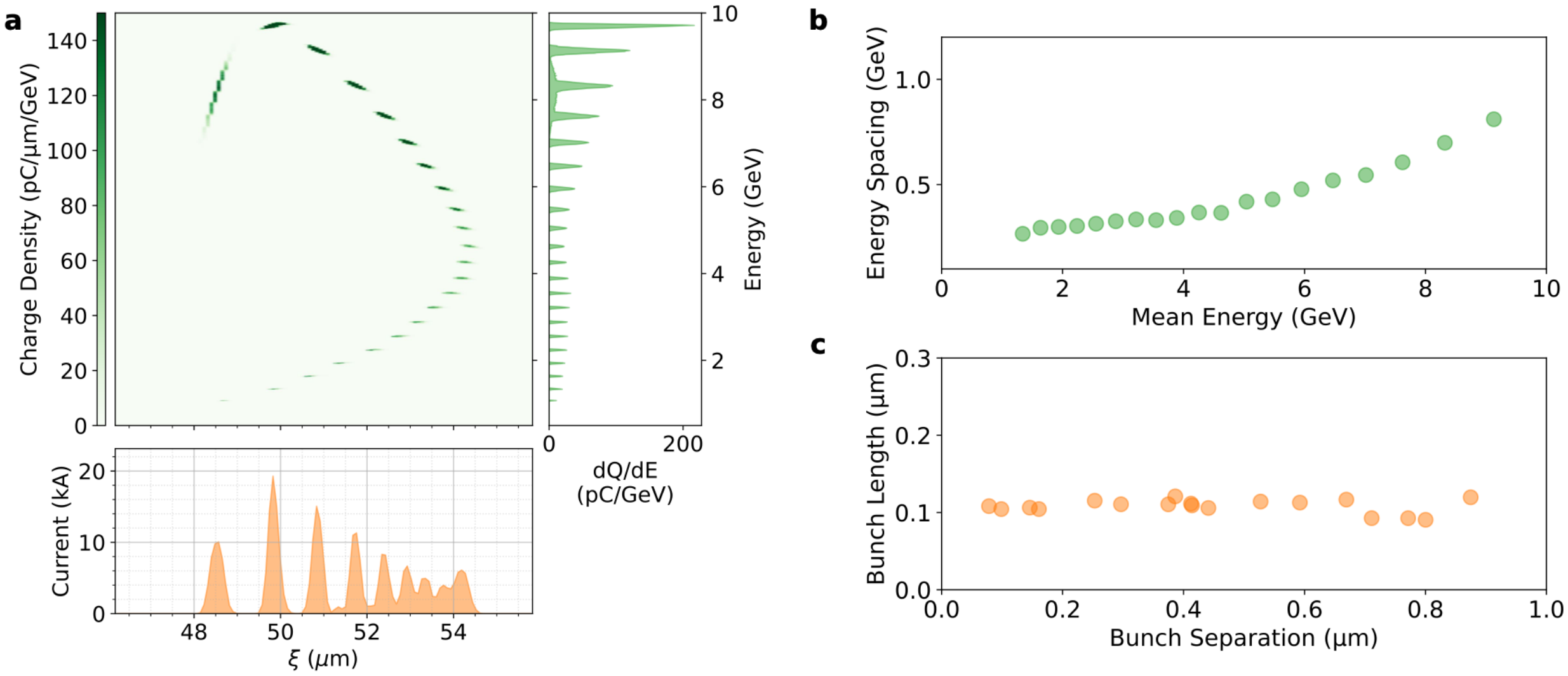

**Extended Data Fig. 3. Effect of beam loading on energy spacing trend.** Simulation with helium density reduced by a factor of 2.5 compared to Fig. 1, resulting in lower charge per bunchlet, smaller peak currents, and weaker beam loading. **a**, Longitudinal phase space (top left), energy spectrum (top right), and current profile (bottom). The comb structure persists with more than ten bunchlets spanning 1-10 GeV. **b**, Energy spacing versus mean energy increases monotonically, in contrast to the parabolic trend observed experimentally (Fig. 3a) and in the baseline simulation (Fig. 3b). **c**, Bunch length versus bunch separation, showing sub-femtosecond durations (~0.1 µm, corresponding to ~0.3 fs).

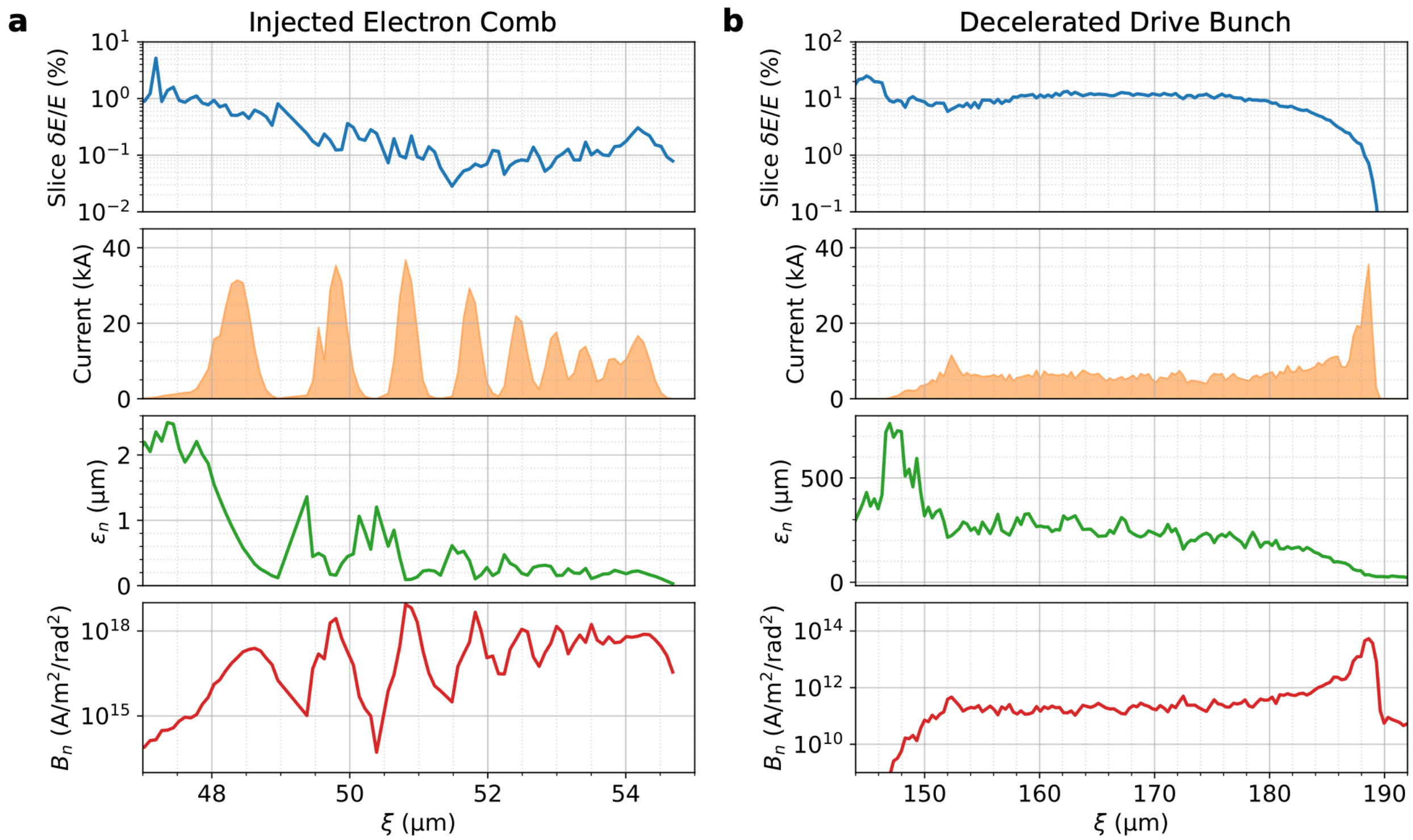

**Extended Data Fig. 4. Simulated slice beam parameters. a**, Slice parameters of the injected electron comb at plasma exit. From top to bottom: rms energy spread, current profile, normalized emittance, and brightness. The injected beam exhibits slice energy spreads as low as 0.1%, normalized emittances down to 0.2 μm, and brightness exceeding $10^{18}$ A/m$^2$/rad$^2$. **b**, Corresponding slice parameters for the decelerated drive beam, showing normalized emittance growth to ~200 μm due to mismatched propagation in the plasma.

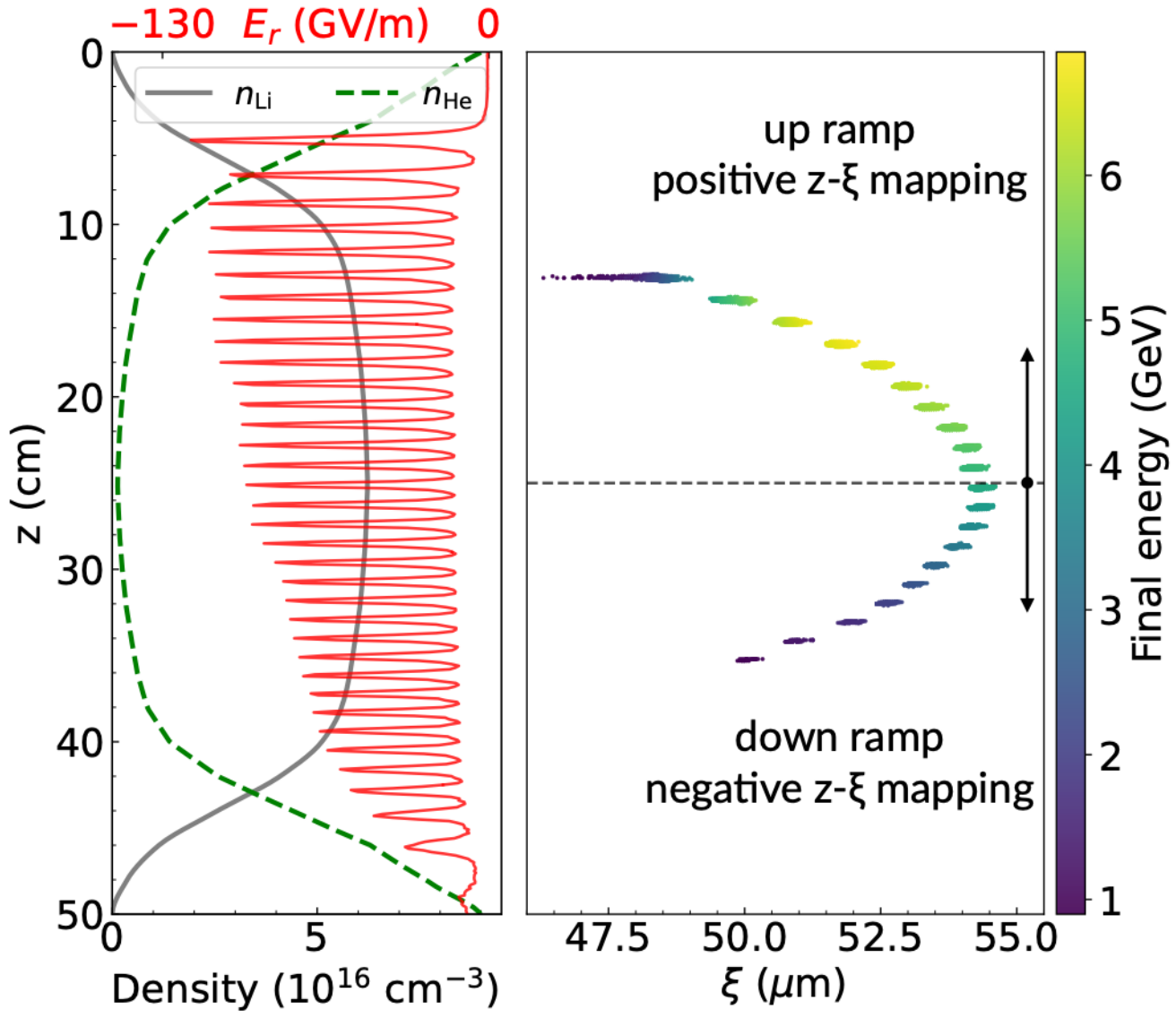

**Extended Data Fig. 5. Density-gradient-mediated phase-space mapping.** Ionization position z versus final co-moving position $\xi$ of injected electrons (from PIC simulation), colored by final energy. Overlaid: lithium plasma density (gray) and transverse electric field $E_r$ of the trailing current spike (red; same as in Fig. 1b) showing periodic betatron pinches where injection occurs. The gentle density up-ramp (z=10-25 cm) produces positive z-$\xi$ mapping: later-ionized electrons are trapped ahead of earlier ones. The down-ramp (z=25-40 cm) reverses this, producing negative mapping. The dashed line marks the density maximum where the mapping reverses and multiple bunchlets converge to similar $\xi$. Electrons ionized over ~20 cm are compressed into a ~6 μm beam, yielding a mapping factor exceeding 30,000.

**Extended Data Movie. Plasma wakefield evolution and electron comb formation.** Animation showing the evolution of the plasma wake, drive beam, and injected electrons as they propagate through the 50-cm-long plasma. Top panel: Lithium (gray) and helium (green) density profiles, with a vertical dashed line indicating the propagation position z. Middle left: Simulation snapshot showing the plasma wake (blue), double-spike drive beam (orange, with current profile in black), ionized helium electrons (cyan), and trapped electron comb (magenta); inset shows detail of the injected bunchlets. Bottom left: On-axis longitudinal electric field Ez. Middle right: Longitudinal phase space of the injected beam, showing charge density versus position $\xi$ and energy, with the projected energy spectrum shown in the right panel and the current profile shown in the panel below. This movie illustrates wake excitation, betatron-induced periodic ionization injection, backward slippage and trapping of helium electrons, and the gradual formation of the electron comb.

**Data availability.** The data that support the findings of this study are available from the corresponding author upon reasonable request.

**Code availability.** The OSIRIS v4.0 code is available from the OSIRIS Consortium upon signing a license agreement.

**Acknowledgements.** The FACET-II E300 plasma wakefield acceleration experiments were built and operated with funding from the U.S. Department of Energy under Contract No. DE-AC02-76SF00515. This work was supported at UCLA by the U.S. Department of Energy through Grant No. DE-SC0010064. Simulations were performed using resources of the National Energy Research Scientific Computing Center (NERSC), a U.S. Department of Energy Office of Science User Facility located at Lawrence Berkeley National Laboratory, operated under Contract No. DE-AC0205CH11231 using NERSC Award HEP-ERCAP-MP113. S.C. was supported by the ANR (g4QED project, Grant No. ANR-23-CE30-0011). C.A.L. was supported by the European Research Council (ERC Grant No. 101116161). O.G.F. was supported by the Research Council of Norway (NFR Grant No. 313770).

**Author contributions.** C.Z. conceived the study. D.S., A.K., B.D.O., C.Z., R.A., O.G.F., and V.L. conducted the experiment. K.A.M. developed the plasma source. D.S., A.K., B.D.O., R.A., S.C., C.E., S.G., C.H., V.L., M.L., and N.M. developed the diagnostics and data acquisition system. C.Z. performed the data analysis and PIC simulations, with help from T.N.D. on the latter. C.E. performed the linac beamline simulations. C.Z. and C.J. wrote the paper, with input from S.C., C.A.L., D.S., T.N.D., W.B.M. and M.J.H. All authors discussed the results and contributed to the final manuscript.